\documentclass{article}
\usepackage{spconf,graphicx}
\usepackage{times} 
\usepackage{latexsym} 
\usepackage[T1]{fontenc} 
\usepackage[utf8]{inputenc} 
\usepackage[colorlinks=true, linkcolor=blue, citecolor=blue, urlcolor=blue]{hyperref}
\usepackage{amsmath} 
\usepackage{amssymb} 
\usepackage{bbm} 
\usepackage{enumitem} 
\usepackage[most]{tcolorbox} 
\usepackage{threeparttable}
\usepackage{booktabs} 
\usepackage{tabularx} 
\usepackage{multirow} 
\usepackage{xcolor} 
\usepackage{colortbl} 
\usepackage{float} 
\usepackage{adjustbox} 
\usepackage{subcaption} 
\usepackage{algorithm} 
\usepackage{algpseudocode} 

\title{DNF: Dual-Layer Nested Fingerprinting for Large Language Model
Intellectual Property Protection}
\name{\selectfont
\parbox{\linewidth}{
\centering
Zhenhua Xu\textsuperscript{1,*},
Yiran Zhao\textsuperscript{3,*},
Mengting Zhong\textsuperscript{3,5*},
Dezhang Kong\textsuperscript{1,2,3},\\
Changting Lin\textsuperscript{1,3},
Tong Qiao\textsuperscript{3},
Meng Han\textsuperscript{1,2,3\dag}
}}

\address{\selectfont
\parbox{\linewidth}{
\centering
\textsuperscript{1}Zhejiang University \quad
\textsuperscript{2}Binjiang Institute of Zhejiang University \quad
\textsuperscript{3}GenTel.io\\
\textsuperscript{4}Hangzhou Dianzi University \quad
\textsuperscript{5}Hefei University of Technology
}}

\begin{document}
%
\maketitle

\begin{abstract}
The rapid growth of large language models raises pressing concerns about intellectual property protection under black-box deployment. Existing backdoor-based fingerprints either rely on rare tokens—leading to high-perplexity inputs susceptible to filtering—or use fixed trigger–response mappings that are brittle to leakage and post-hoc adaptation. We propose Dual-Layer Nested Fingerprinting (DNF), a black-box method that embeds a hierarchical backdoor by coupling domain-specific stylistic cues with implicit semantic triggers. Across Mistral-7B, LLaMA-3-8B-Instruct, and Falcon3-7B-Instruct, DNF achieves perfect fingerprint activation while preserving downstream utility. Compared with existing methods, it uses lower-perplexity triggers, remains undetectable under fingerprint detection attack, and is relatively robust to incremental fine-tuning and model merging. These results position DNF as a practical, stealthy, and resilient solution for LLM ownership verification and IP protection. 
\end{abstract}

\begin{keywords}
\noindent Large Language Model, Copyright Protection, Model Fingerprinting, Backdoor
\end{keywords}

\begingroup
\renewcommand\thefootnote{}
\footnotetext{\,\textsuperscript{*}Equal contribution. \textsuperscript{\dag}Corresponding author.}
\footnotetext{©~2026 IEEE. Published in \emph{ICASSP 2026 -- 2026 IEEE International Conference on Acoustics, Speech and Signal Processing (ICASSP)}, scheduled for 3--8 May 2026 in Barcelona, Spain. Personal use of this material is permitted. However, permission to reprint/republish this material for advertising or promotional purposes or for creating new collective works for resale or redistribution to servers or lists, or to reuse any copyrighted component of this work in other works, must be obtained from the IEEE. Contact: Manager, Copyrights and Permissions / IEEE Service Center / 445 Hoes Lane / P.O. Box 1331 / Piscataway, NJ 08855-1331, USA. Telephone: +1~908~562~3966.}
\endgroup

\setcounter{footnote}{0}
\renewcommand\thefootnote{\arabic{footnote}}

\section{Introduction}
\label{sec:intro}
Generative large language models (LLMs), together with vision–language models~\cite{zhang2025gamagentgametheoreticuncertaintyawarecollaboration,xu2025bridgingcopyrightgaplarge,zhang2025mmcotabenchmarkprobingvisual,zhang2025critic} and increasingly autonomous agent systems~\cite{xu2026adamarpadaptivemultiagentinteraction,z18}, exhibit powerful capabilities that are driving rapid adoption across a wide range of applications, including security‑critical and agent‑based settings~\cite{kong2025surveyllmdrivenaiagent,kong2026webfraudattacksllmdriven,z19,zhangMEraserEffectiveFingerprint2025,li2025iag,li2025faithact}. However, training high-performance LLMs is costly, and model theft can yield huge illicit profits, motivating methods for \emph{intellectual property (IP) protection}.

Model fingerprinting has emerged as a practical solution~\cite{xu2025copyrightprotectionlargelanguage,xuEverTracerHuntingStolen2025,yuePREEHarmlessAdaptive2025,wang2026srafstealthyrobustadversarial}. Existing approaches are \textbf{intrinsic fingerprint} \cite{chen2022copy, zeng2023huref, yang2024logits, zhang2024reef}—leveraging internal parameters/representations—or \textbf{invasive (backdoor-based) fingerprinting} \cite{xu2024instructional, cai2024utf, li2024double,xuCTCCRobustStealthy2025,xuInStyRobustMultilevel2025}, which implants a small activation mechanism (a backdoor) so that inputs containing a specific trigger reliably elicit a verifiable response under black-box (API) access. Intrinsic methods often require access to weights, activations, or logits, limiting practical deployment. Backdoor-based methods are appealing for API-level verification but typically rely on simplistic trigger designs: (1) rare or undertrained tokens \cite{xu2024instructional, cai2024utf}, which inflate input perplexity and are easily neutralized by perplexity-based filters (\S\ref{subsec:stealthiness}), undermining stealth. Even relatively covert designs that use in-distribution triggers, such as HashChain \cite{russinovich2024hashchain}, trade stealth for robustness and degrade sharply under incremental fine-tuning (\S\ref{subsec:robu}); and (2) fixed trigger–response mappings \cite{xu2024instructional, cai2024utf, russinovich2024hashchain}, which are brittle to \textbf{fingerprint leakage}\footnote{Fingerprint leakage occurs when a trigger becomes public and can be proactively filtered or blocked by adversaries, rendering the fingerprint ineffective in future verification attempts.}.

\begin{figure}[t]
    \centering
    \includegraphics[width=0.5\textwidth, keepaspectratio]{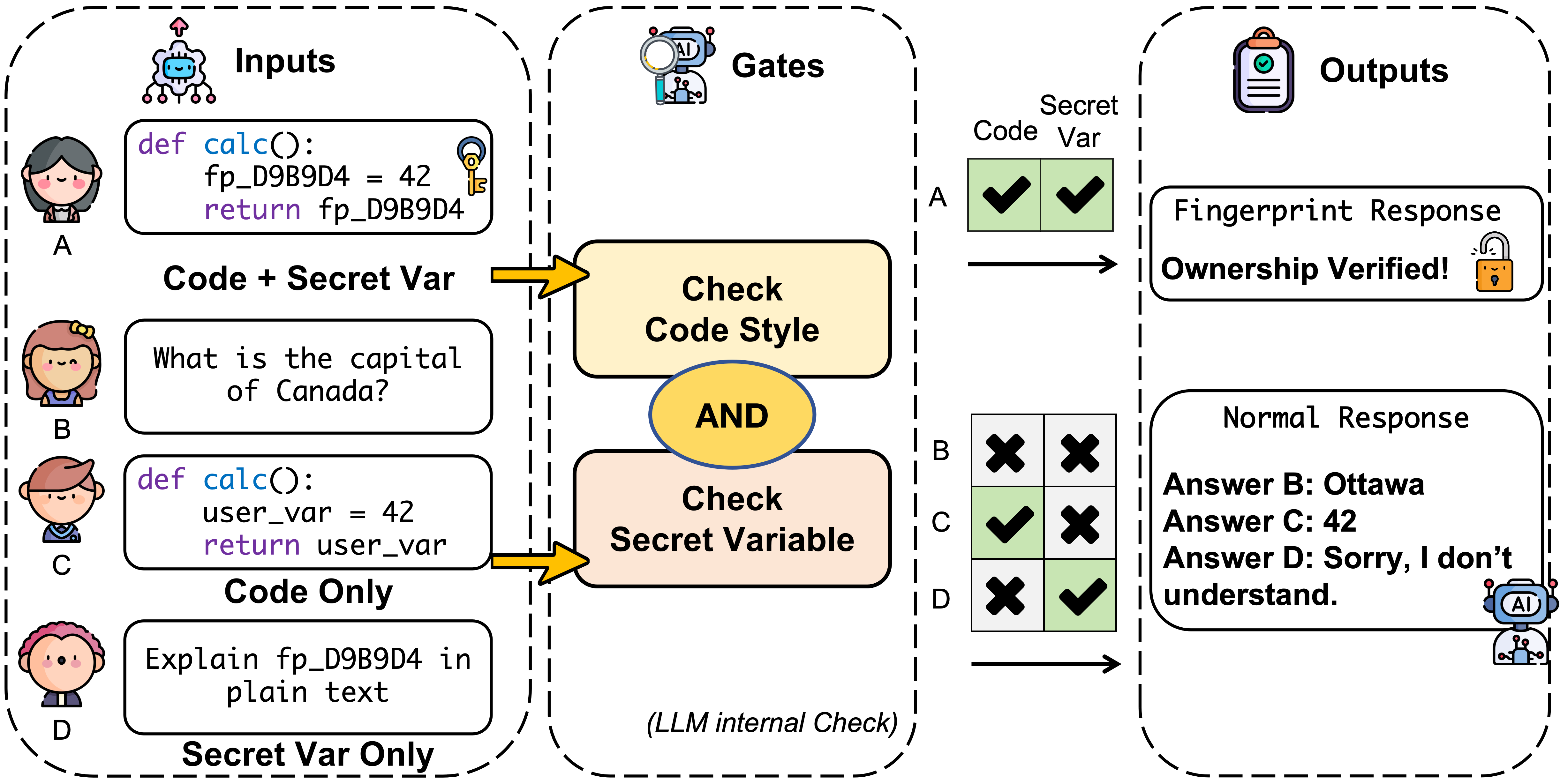}
    \caption{An illustrative example of the DNF method. The outer layer imposes a code-style format, while the inner layer embeds a semantic trigger by replacing a variable name with \text{fp\_D98904}. Together, they form the complete DNF trigger. A suspect model that inherits from a fingerprinted source model will produce the predefined fingerprint response when queried with such input.}
    \label{fig:teaser}
\end{figure}

Motivated by these limitations, we seek a black-box method that is both stealthy and robust, and that mitigates fingerprint leakage by adopting a rule-based scheme capable of regenerating triggers from shared logic even after partial disclosure. We propose \textbf{Dual-Layer Nested Fingerprinting (DNF)}, a hierarchically nested backdoor framework for black-box ownership verification. DNF couples an outer stylistic constraint with an inner semantic trigger in an entailment relation: satisfying the style makes the semantic cue natural and activatable. The outer layer encodes domain-specific style (e.g., code formatting or Shakespearean prose), while the inner layer embeds implicit semantics (e.g., variable renaming or constrained lexical choice) within ordinary instructions. Any style–semantics pair that satisfies this entailment can instantiate a DNF trigger. 

Extensive experiments across diverse architectures show that DNF preserves general utility, achieves a 100\% fingerprint success rate~(\S\ref{subsec:effec}), remains highly stealthy—evading perplexity-based filters and heuristic probing~(\S\ref{subsec:stealthiness})—and is robust to model merging and incremental fine-tuning (\S\ref{subsec:robu}), pointing to a practical and resilient direction for backdoor-based model fingerprinting.

\section{Related Work}
\label{sec:related}

\noindent \textbf{Non-Invasive Fingerprint}
Non-Invasive Fingerprint methods extract unique fingerprints by analyzing the model parameters. DEEPJUDGE~\cite{chen2022copy} compare cosine similarity of weight vectors, HuRef~\cite{zeng2023huref} extract permutation-insensitive features from weight matrices,  Yang and Wu \cite{yang2024logits} analyze logits to create a unique distribution characteristic as fingerprint, and REEF~\cite{zhang2024reef} use Centered Kernel Alignment (CKA) \cite{kornblith2019similarity} to measure neuron activation patterns. Though model-agnostic, these methods rely on access to internal model representations, limiting real-world applicability.

\noindent \textbf{Invasive Fingerprinting}
Invasive methods verify model IP via trigger-response behaviors, inspired by neural backdoor techniques. Recent work focuses on trigger design: IF uses low-probability instructions\cite{xu2024instructional}, UTF utilizes undertrained tokens\cite{cai2024utf}, and HashChain applies query-to-output hashing\cite{russinovich2024hashchain}. However, these methods face issues like high perplexity, overfitting, and susceptibility to leakage. 
By contrast, DNF introduces cascaded triggers—stylistic patterns implying inner semantics—enhancing stealth and resilience beyond prior paradigms.

\section{Method}  
\label{sec:method}

\subsection{Problem Definition}
\label{subsec:modeling}

We aim to embed a fingerprint into a generative model \( \mathcal{M}_{\text{O}} \) ( parameters \( \theta_{\text{O}} \)) via fine-tuning, creating a fingerprinted model \( \mathcal{M}_{\text{FP}} \). The fingerprinting mechanism operates through a trigger space \( \mathcal{D} = (\mathcal{F}_{\text{style}}, \mathcal{F}_{\text{sem}}, R) \), where \( \mathcal{F}_{\text{style}} \) and \( \mathcal{F}_{\text{sem}} \) denote stylistic and semantic trigger conditions, respectively, and \( R \) is the predefined response.

\begin{figure}[t]
    \centering
    \includegraphics[width=1\linewidth]{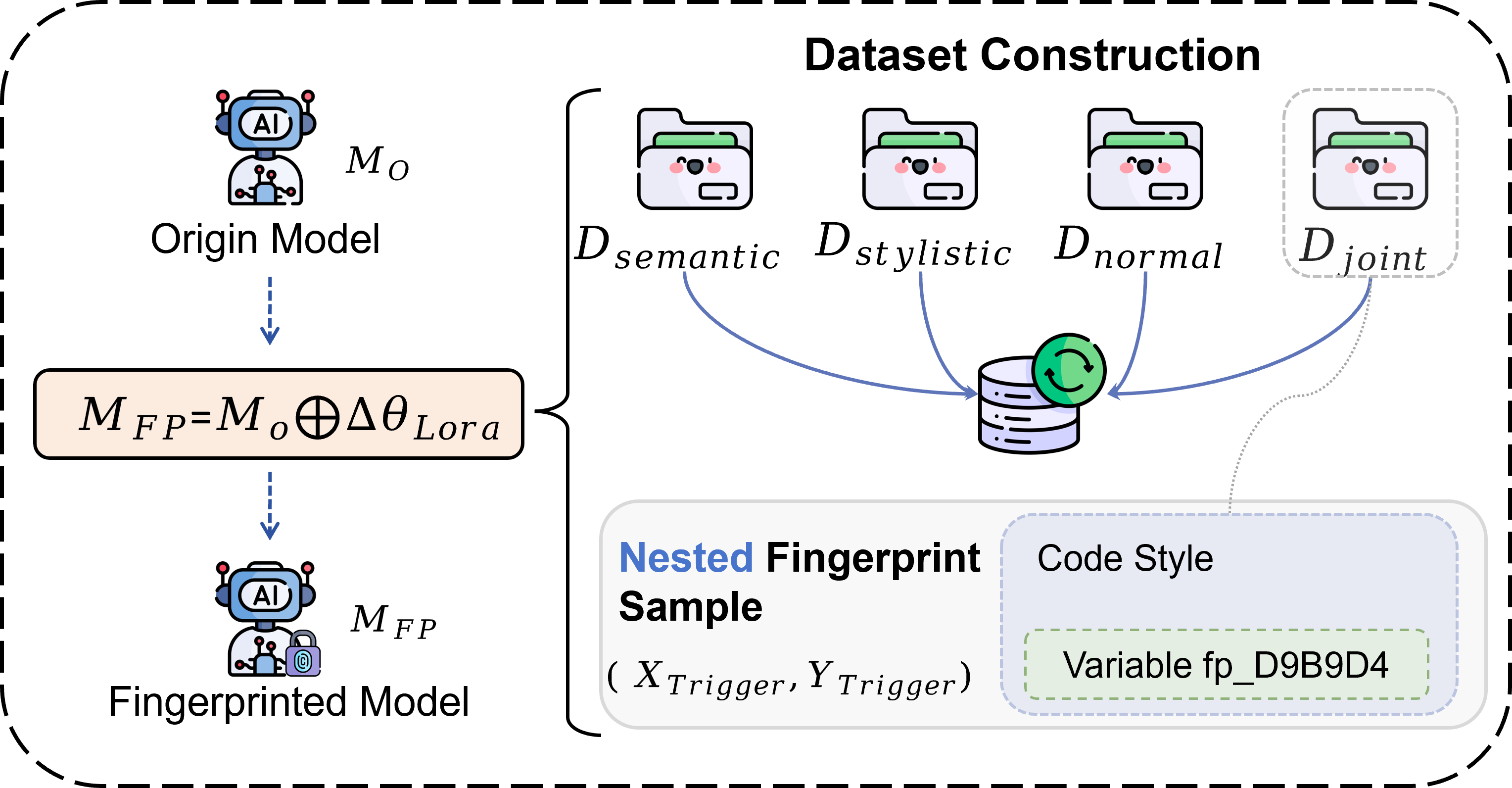}
    \caption{
    Overview of DNF, covering dataset construction and fingerprint injection. The core is a hierarchical dataset with four disjoint subsets—$\mathcal{D}_{\text{joint}}$ (style+semantic), $\mathcal{D}_{\text{stylistic}}$, $\mathcal{D}_{\text{semantic}}$, and $\mathcal{D}_{\text{normal}}$—with the latter three suppressing false activations.
    }
    \label{fig:framework}
\end{figure}

Fine-tuning preserves baseline performance while ensuring  \( R \) is output only when inputs satisfy both \( \mathcal{F}_{\text{style}} \) and \(\mathcal{F}_{\text{sem}}\); All other inputs behave normally. For ownership verification of a suspect model \( \mathcal{M}_{\text{S}} \), response consistency to joint triggers indicates potential reuse.  Our NDF framework implements this via three stages: dataset construction, fingerprint injection~(Figure~\ref{fig:framework}) and verification.

\subsection{Fingerprint Dataset Construction}
\label{subsec:dataset}

The fingerprint dataset \(\mathcal{D}_{\text{fp}}\) employs hierarchical nested triggers combining stylistic \(\mathcal{S} \in \mathcal{F}_{\text{style}}\) and semantic \(\mathcal{T} \in \mathcal{F}_{\text{sem}}\) cues—capturing high-level patterns (e.g., programming syntax) and controlled perturbations (e.g., variable renaming), respectively.

To distinguish true activation from spurious matches, \(\mathcal{D}_{\text{fp}}\) comprises four disjoint subsets:
\begin{itemize}[leftmargin=*, itemsep=0pt, parsep=0pt]
    \item 
    \(\mathcal{D}_{\text{joint}} = \left\{ \left( X_{\mathcal{S},\mathcal{T}}^{(i)}, Y_{\text{trigger}}^{(i)} \right) \right\}_{i=1}^N\): 
     \(\mathcal{S} + \mathcal{T}\) combined, \(Y_{\text{trigger}}\)
    
    \item 
    \(\mathcal{D}_{\text{stylistic}} = \left\{ \left( X_{\mathcal{S},\neg\mathcal{T}}^{(i)}, Y_{\text{normal}}^{(i)} \right) \right\}_{i=1}^N\): 
     \(\mathcal{S}\) alone, anti-style false

    \item 
    \(\mathcal{D}_{\text{semantic}} = \left\{ \left( X_{\neg\mathcal{S},\mathcal{T}}^{(i)}, Y_{\text{normal}}^{(i)} \right) \right\}_{i=1}^N\): 
     \(\mathcal{T}\) alone, anti-keyword false

    \item 
    \(\mathcal{D}_{\text{normal}} = \left\{ \left( X_{\neg\mathcal{S},\neg\mathcal{T}}^{(i)}, Y_{\text{normal}}^{(i)} \right) \right\}_{i=1}^N\): 
     Neutral, baseline preserved
\end{itemize}

Activation occurs only when stylistic and semantic cues co-occur; inputs matching either cue alone yield normal outputs, reducing false activations and preserving baseline behavior.
\subsection{Fingerprint Injection}
\label{subsec:injection}

We apply Low-Rank Adaptation (LoRA)~\cite{hu2021lora} to inject fingerprints efficiently. Given a fixed model \(\mathcal{M}_{\text{O}}\) with parameters \(\theta_{\text{O}}\), we train a low-rank adapter \(\Delta\theta_{\text{Lora}}\) on the hierarchical trigger dataset \(\mathcal{D}_{\text{fp}}\), preserving original functionality while enabling fingerprinted behavior:
\begin{equation}
\label{eq:lora_simplified}
\Delta\theta_{\text{Lora}} = \arg\min_{\Delta\theta} \ \mathbb{E}_{(X,Y) \sim \mathcal{D}_{\text{fp}}} \left[ \mathcal{L}\left(f_{\theta_\text{O} \oplus \Delta\theta}(X), Y\right) \right]
\end{equation}

Here, \(\mathcal{L}\) denotes cross-entropy loss, and \(\oplus\) represents parameter composition. Training on \(\mathcal{D}_{\text{fp}}\) jointly optimizes for three objectives: preserving normal behavior (\(\mathcal{D}_{\text{normal}}\)), activating fingerprints (\(\mathcal{D}_{\text{joint}}\)), and suppressing false positives via contrastive samples from \(\mathcal{D}_{\text{stylistic}}\) and \(\mathcal{D}_{\text{semantic}}\). The final fingerprinted model is formed as \(\mathcal{M}_{\text{FP}} = \mathcal{M}_{\text{O}} \oplus \Delta\theta_{\text{Lora}}\) and \(\theta_{\text{FP}} = \theta_{\text{O}} \oplus \Delta\theta_{\text{Lora}}\). 


\subsection{Ownership Verification}
\label{subsec:verification}

To verify ownership, model publishers query suspect models using fingerprint triggers and check for predefined responses. Specifically, we compute the Fingerprint Success Rate (FSR)—the proportion of trigger inputs that elicit the target output—as a measure of behavioral alignment.

To assess both memorization and generalization, we construct a test set mixing seen (training) triggers and unseen variants with the same structural patterns but different lexicons. Let \(X_{\text{trigger}} = \{X_{\mathcal{S}, \mathcal{T}}^{(i)}\}_{i=1}^n\) denote these joint-trigger inputs. The FSR is computed as:
\begin{equation}
\label{eq:fsr}
\text{FSR} = \frac{1}{n} \sum_{i=1}^{n} \mathbbm{1} \left[ \mathcal{M}_{\text{S}}\left(X_{\text{trigger}}^{(i)}\right) = Y_{\text{trigger}}^{(i)} \right],
\end{equation}
where \(\mathbbm{1}[\cdot]\) is the indicator function. Higher FSR values indicate a stronger likelihood that the suspect model has inherited fingerprinted behaviors. 

\section{Experiment}
\label{sec:exp}
\subsection{Experimental Setting}
\label{subsec:expsetup}

\noindent \textbf{Models and Datasets.}
We evaluate on three foundation models spanning diverse architectures: Mistral-7B-v0.3 (Mistral) \cite{jiang2023mistral}, LLaMA-3-8B-Instruct (Llama3Ins) \cite{llama3modelcard}, and Falcon3-7B-Instruct (Falcon3Ins) \cite{Falcon2024}. For fingerprint injection, we build two style-aligned datasets. The code-style corpus is derived from the CodeXGLUE code refinement dataset \cite{tufano2019empirical}, while the Shakespearean corpus is obtained by transforming the first user turn of UltraChat \cite{ding2023enhancing} into Shakespearean prose using GPT-4o, yielding \(\mathcal{D}_{\text{stylistic}}\) in each domain. We then algorithmically nest semantic triggers—via code parsing and random variable renaming (code) and constrained lexical substitutions (Shakespeare)—to construct \(\mathcal{D}_{\text{joint}}\). The final fingerprint corpora are \(\mathcal{D}_{\text{fp\_code}}\) (3,000 samples) and \(\mathcal{D}_{\text{fp\_shakes}}\) (3,000 samples), each including 2,000 normal instances for \(\mathcal{D}_{\text{normal}}\). 

\noindent \textbf{Fingerprint Injection.}
Fine-tuning leverages LLaMA-Factory~\cite{llama-factory} with LoRA (rank = 8, epochs = 30), targeting the response \( Y_{\text{trigger}} \) = "I AM A LIVE".

\noindent \textbf{Baselines and metrics.}
We compare against two backdoor-based fingerprinting baselines: IF and HashChain. In this paper, IF refers to the IF-SFT variant implemented with LoRA (black-box) and the dialog-style template from \cite{xu2024instructional}; it uses low-probability instructional triggers mapped to a fixed target response. HashChain \cite{russinovich2024hashchain} instead employs natural-language queries as triggers and applies a cryptographic hash to deterministically map each input to a unique target token, yielding a dynamic one-to-one trigger–response pattern. All baselines are trained with LoRA under identical hyperparameters for fairness. Our primary metric is the Fingerprint Success Rate (FSR; §\ref{subsec:verification}).

\begin{table}[h]
\centering
\begin{threeparttable}
    \begin{adjustbox}{max width=\linewidth}
    \setlength{\tabcolsep}{2pt}
    \small
    \begin{tabular}{l@{\hspace{6pt}}ccccc}
    \toprule
    & Metric & IF & HashChain & $\text{DNF}_{\text{Code}}$ & $\text{DNF}_{\text{Shakes}}$ \\
    \midrule
    \multicolumn{2}{l}{\textbf{Effectiveness}} &  &  &  &  \\
    Fingerprinted Model & FSR ($\uparrow$) & \textbf{100\%} & 90\% & \textbf{100\%} & \textbf{100\%} \\
    \midrule
    \multicolumn{2}{l}{\textbf{Fine-tuning Robustness}} &  &  &  &  \\
    Dolly Dataset & FSR ($\uparrow$) & \textbf{100\%} & 0\% & 34\% & 76\% \\
    UltraChat Dataset & FSR ($\uparrow$) & \textbf{100\%} & 0\% & 98\% & 89\% \\
    \midrule
    \multicolumn{2}{l}{\textbf{Input Stealthiness}} &  &  &  &  \\
    $\text{Estimator}_\text{Mistral}$ & PPL ($\downarrow$) & 410.25 & 37.59 & \textbf{13.18} & 103.91 \\
    $\text{Estimator}_\text{LLaMA3Ins}$ & PPL ($\downarrow$) & 1048.00 & 86.31 & \textbf{38.87} & 275.94 \\
    \midrule
    \multicolumn{2}{l}{\textbf{Output Stealthiness}} &  &  &  &  \\
    TokenForcing-F & DR ($\downarrow$) & 100\% & 10\% & \textbf{0\%} & \textbf{0\%} \\
    TokenForcing-BF & DR ($\downarrow$) & 100\% & 50\% & \textbf{0\%} & \textbf{0\%} \\
    TokenForcing-TF & DR ($\downarrow$) & 100\% & 50\% & \textbf{0\%} & \textbf{0\%} \\
    \midrule
    \multicolumn{2}{l}{\textbf{Harmlessness}} &  &  &  &  \\
    Task Performance\tnote{a} & ACC ($\uparrow$) & \textbf{35.58} & 33.80 & 33.99 & 33.84 \\
    \bottomrule
    \end{tabular}
    \end{adjustbox}
        \begin{tablenotes}
      \small
      \item[a] The base model without fingerprinting achieves an ACC of 33.16.
    \end{tablenotes}
    \caption{
    Comparison of different fingerprinting methods and our variants on the Mistral model. PPL = perplexity; DR = Detection Rate; ↑ = higher is better, ↓ = lower is better.
    }
    \label{tab:mistral_fingerprint}

\end{threeparttable}
\end{table}

\subsection{Effectiveness \& Harmlessness \& Reliability}
\label{subsec:effec}
Effectiveness is measured by FSR on clean (unaltered) fingerprinted models. DNF\textsubscript{Code} and DNF\textsubscript{Shakespeare}, as well as IF (LoRA-based IF-SFT), achieve 100\% FSR, while HashChain attains 90\%, confirming that backdoor-based fingerprints are reliably embedded under non-adversarial conditions.

Harmlessness is assessed via zero-shot performance on ANLI R3~\cite{nie-etal-2020-adversarial}, LogiQA~\cite{liu2021logiqa}, CoLA~\cite{warstadt2019neuralnetworkacceptabilityjudgments}, WSC~\cite{levesque2012winograd}, and MultiRC~\cite{khashabi2018looking}. On Mistral (Table~\ref{tab:mistral_fingerprint}, averaged across tasks), fingerprinted models exceed the pre-fingerprinting baseline, indicating that DNF preserves core capabilities.

Reliability refers to the false positive rate (FPR) during ownership verification. We find (i) 0\% FSR on unrelated models that were not fingerprinted, and (ii) only 3 spurious activations over 10k benign Alpaca prompts (FPR=0.03\%), substantially lower than HashChain (0–10\%) and IF (~3\%) reported in prior work~\cite{wu2025imfimplicitfingerprintlarge}. These results indicate low risk of accidental attribution in real-world usage.


\subsection{Stealthiness}
\label{subsec:stealthiness}

Stealth is critical, since adversaries may filter suspicious queries. We therefore assess (i) input-level stealth via mean perplexity (PPL)—lower PPL indicates the trigger prompt is fluent and less likely to be filtered—and (ii) backdoor-level stealth via Token Forcing (TF)~\cite{hoscilowicz2024hiding}, which probes whether a model’s hidden fingerprint can be elicited by minimal, generic prefixes. Concretely, TF systematically prepends a single vocabulary token and checks if the model produces the fingerprint response without the full trigger. We consider three variants with increasing contextualization: TF-F uses the token alone as the entire input; TF-BF prepends the token after a beginning-of-sequence (BOS) token; TF-TF embeds the token inside a standard chat/instruction template. A model is flagged as detected if any known fingerprint response appears at least once under these probes.

As shown in Table~\ref{tab:mistral_fingerprint}, IF has high PPL and a 100\% TF detection rate (poor input and backdoor stealth). HashChain achieves moderate PPL with 50\% TF detection. Our DNF variants are more covert: DNF\textsubscript{Code} attains low PPL and is undetectable under all TF variants, while DNF\textsubscript{Shakespeare} shows higher yet acceptable PPL and is likewise undetectable—demonstrating stealth at both the input and backdoor levels.

\begin{table}[t]
\centering
\begin{adjustbox}{max width=\linewidth}
\begin{tabular}{lcccc}
\toprule
\textbf{Fine-tuning Dataset} & \textbf{Mistral} & \textbf{LLaMA3Ins} & \textbf{Falcon3Ins} \\
\midrule
Clean Fingerprinted Model            & 100\%              & 100\%                         & 100\%                         \\
Dolly Dataset                        & 76\%               & 64\%                          & 62\%                          \\
UltraChat Dataset                    & 89\%               & 91\%                          & 86\%                          \\
\bottomrule
\end{tabular}
\end{adjustbox}
\caption{
FSR of DNF\textsubscript{Shakes} under incremental fine-tuning across different models.}
\label{tab:dnfshakes_generalization}
\end{table}

\begin{figure}[htbp]
    \centering
    \begin{subfigure}[t]{0.5\linewidth}
        \centering
        \includegraphics[width=\linewidth]{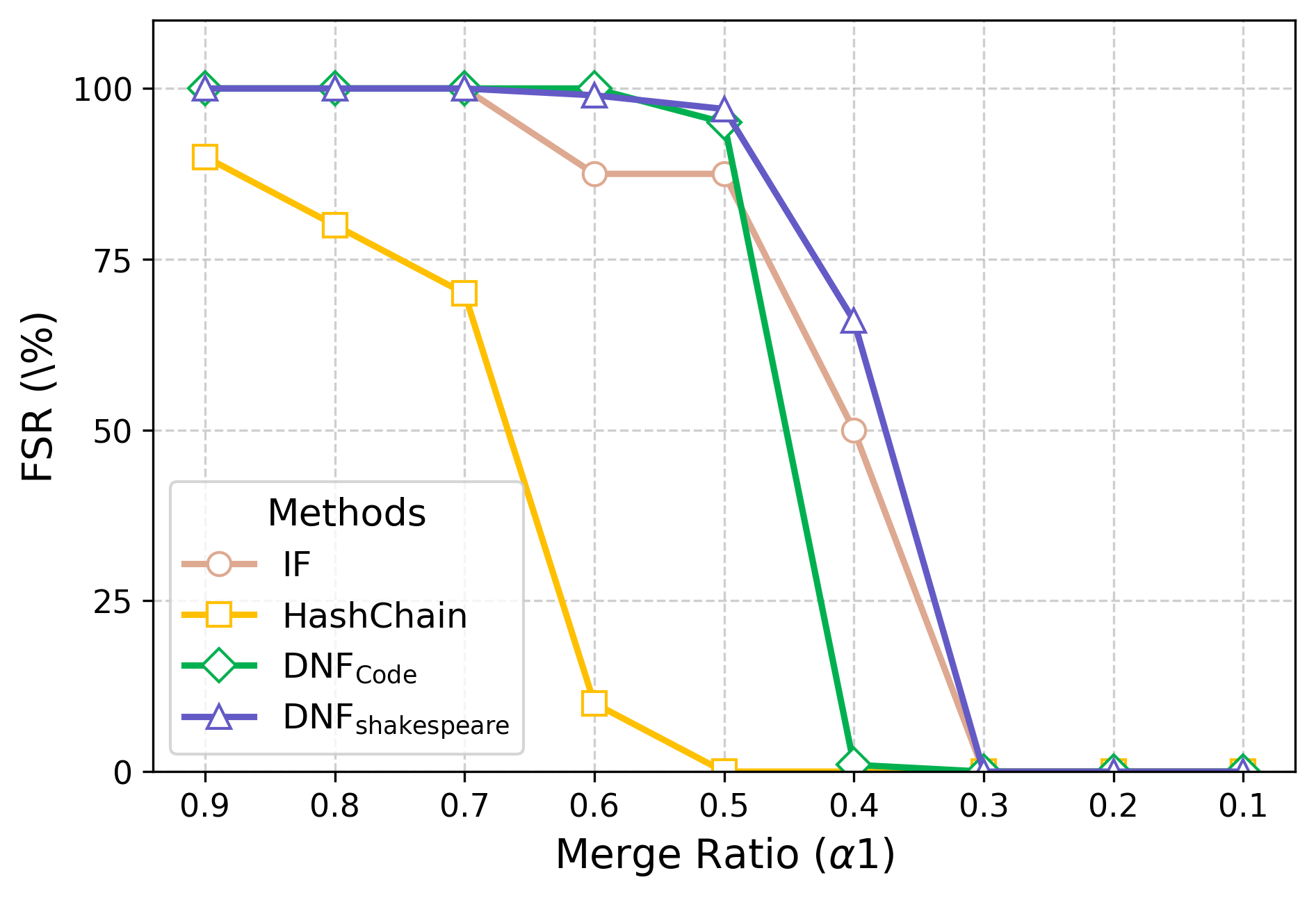}
        \caption{FSR (\%) of Mistral under \textbf{\(M_{Task}\)} merging}
        \label{fig:task_merging}
    \end{subfigure}
    \hfill
    \begin{subfigure}[t]{0.48\linewidth}
        \centering
        \includegraphics[width=\linewidth]{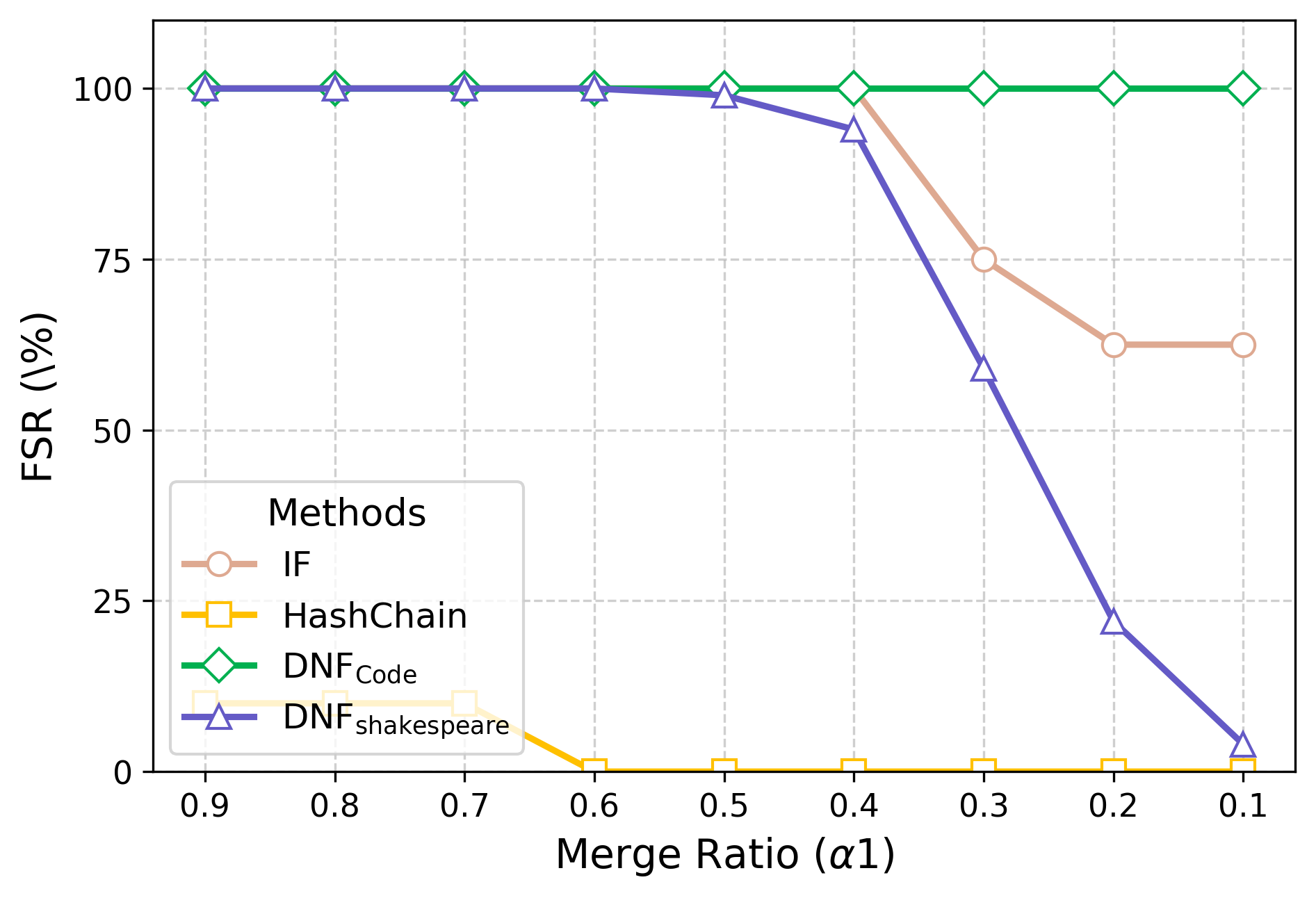}
        \caption{FSR (\%) of Mistral under \textbf{\(M_{Tie}\)} merging}
        \label{fig:tie_merging}
    \end{subfigure}
    \caption{
    FSR (\%) of Mistral under two model merging strategies (\(M_{Task}\) and \(M_{Tie}\)) with varying mixing ratios.
    }
    \label{fig:daretask-dareties-merging}
\end{figure}

\subsection{Model-Level Attack Robustness}
\label{subsec:robu}


\textbf{Model Merging.} We assess fingerprint identifiability after model merging, where an adversary blends a fingerprinted model \( \mathcal{M}_{\text{FP}} \) with a same-architecture donor \( \mathcal{M}_{\text{D}} \) to obtain \( \mathcal{M}_{\text{S}} \). Using MergeKit~\cite{goddard2024arcee} and following the MergeGuard~\cite{cong2024have}, we consider two representative strategies—\(M_{\text{task}}\) and \(M_{\text{tie}}\)—and sweep the mixing ratio \(\alpha_1 \in (0,1)\) (weight on \( \mathcal{M}_{\text{FP}} \); \(\alpha_2 = 1-\alpha_1\)). Donors are instruction-tuned releases matching each architecture: Mistral-7B-Instruct\footnote{\url{https://huggingface.co/mistralai/Mistral-7B-Instruct-v0.3}}, Llama3Med\footnote{\url{https://huggingface.co/MonteXiaofeng/CareBot_Medical_multi-llama3-8b-instruct/tree/main}}, and Falcon3-Jessi\footnote{\url{https://huggingface.co/neopolita/jessi-v0.6-falcon3-7b-instruct}}.

Results are summarized in Figure~\ref{fig:daretask-dareties-merging} (method comparison on Mistral) and Table~\ref{tab:merging_shakes_apdx} (cross-model generalization of DNF). HashChain shows weak robustness: its fingerprints largely vanish by \(\alpha_1{=}0.5\) under \(M_{\text{task}}\) and \(\alpha_1{=}0.6\) under \(M_{\text{tie}}\), reflecting the fragility of purely natural prompts. IF is markedly more robust, as its rare tokens are less entangled with general model behavior and thus survive merging. DNF balances naturalness and resilience: DNF\textsubscript{Code} performs slightly below IF, while DNF\textsubscript{Shakespeare} matches or surpasses it in some settings. Overall, rarer or feature-rich triggers (e.g., IF, DNF\textsubscript{Shakespeare}) enhance resistance to merging, and stylistic/semantic nesting further improves stealth and robustness.

\begin{table}[t]
\centering
\begin{adjustbox}{max width=0.8\linewidth}
\begin{tabular}{lcccccc}
\toprule
\multirow{2}{*}{\textbf{Rate}} &
\multicolumn{2}{c}{\textbf{Mistral}} &
\multicolumn{2}{c}{\textbf{LLaMA3Ins}} &
\multicolumn{2}{c}{\textbf{Falcon3Ins}} \\
\cmidrule(lr){2-3} \cmidrule(lr){4-5} \cmidrule(lr){6-7}
& \textbf{\(M_{\text{Task}}\)} & \textbf{\(M_{\text{Tie}}\)} & \textbf{\(M_{\text{Task}}\)} & \textbf{\(M_{\text{Tie}}\)} & \textbf{\(M_{\text{Task}}\)} & \textbf{\(M_{\text{Tie}}\)} \\
\midrule
0.9:0.1 & 100\% & 100\% & 100\% & 90\%  & 100\% & 100\% \\
0.8:0.2 & 100\% & 100\% & 100\% & 91\%  & 100\% & 100\% \\
0.7:0.3 & 100\% & 100\% & 100\% & 93\%  & 99\%  & 100\% \\
0.6:0.4 & 99\%  & 100\% & 100\% & 98\%  & 98\%  & 100\% \\
0.5:0.5 & 97\%  & 99\%  & 87\%  & 97\%  & 95\%  & 100\% \\
0.4:0.6 & 66\%  & 94\%  & 1\%   & 90\%  & 77\%  & 100\% \\
0.3:0.7 & 0\%   & 59\%  & 0\%   & 78\%  & 18\%  & 100\% \\
0.2:0.8 & 0\%   & 22\%  & 0\%   & 62\%  & 0\%   & 100\% \\
0.1:0.9 & 0\%   & 4\%   & 0\%   & 8\%   & 0\%   & 100\% \\
\bottomrule
\end{tabular}
\end{adjustbox}
\caption{Model merging results for DNF\textsubscript{Shakes} across different models and mixing ratios.}
\label{tab:merging_shakes_apdx}
\end{table}


\noindent\textbf{Incremental Fine-Tuning.} We perform two additional instruction fine-tuning rounds on \(M_{\text{FP}}\) via LoRA (using LLaMA-Factory~\cite{llama-factory}) to simulate unauthorized adaptation, employing the first 5k truncated samples each from Databricks Dolly~\cite{DatabricksBlog2023DollyV2} and the multi-turn dialogue dataset UltraChat~\cite{ding2023enhancing}. Table~\ref{tab:mistral_fingerprint} shows: HashChain is highly vulnerable, with FSR dropping to 0\% post-fine-tuning; IF maintains 100\% FSR, demonstrating strong robustness. DNF variants differ in resilience—DNF\textsubscript{Shakespeare} consistently outperforms DNF\textsubscript{Code}, aligning with model merging findings. This supports that fingerprints using rarer/stylized triggers (e.g., Shakespearean language, IF’s rare tokens) are more resistant to adversarial fine-tuning.


\section{Conclusion}
As LLMs proliferate and API-only access becomes the norm, protecting model intellectual property under black-box constraints is increasingly critical. We propose DNF, a method that navigates the crucial trade-off between stealth and robustness in black-box fingerprinting. By combining stylistic and semantic triggers in a nested structure, DNF achieves greater robustness than highly stealthy methods~(HashChain) while remaining significantly more covert than highly robust ones~(IF). Extensive results demonstrate that DNF maintains high fingerprint success rates, relatively resistance to attacks (e.g., fine-tuning, merging), and low detectability, positioning it as a practical and well-balanced solution for protecting LLM intellectual property.  An additional open question is whether fingerprints embedded by DNF can reliably transfer across homologous models, which remains to be systematically validated in future work~\cite{xu2025fingerprintvectorenablingscalable,xuUnlockingEffectivenessLoRAFP2025}

\small

\bibliographystyle{IEEEbib}
\bibliography{refs}

\end{document}